\documentclass[11pt]{article}
\usepackage{amsmath,amssymb,amsthm, latexsym,amsfonts}
\usepackage[dvips]{graphics,epsfig}

\title{Bias correction for estimators of the extremal index}
\author{Holger Drees\footnote{University of Hamburg, Department of Mathematics,
SPST, Bundesstr.\ 55, 20146 Hamburg, Germany; email:
holger.drees@uni-hamburg.de}}

\newcommand{\R}{\mathbb{R}}
\newcommand{\N}{\mathbb{N}}
\newcommand{\B}{\mathbb{B}}

\newcommand{\BB}{\mathcal{B}}

\newtheorem{theorem}{Theorem}[section]
\newtheorem{corollary}[theorem]{Corollary}
\newtheorem{remark}[theorem]{Remark}
\newtheorem{example}[theorem]{Example}

\numberwithin{equation}{section}

\newenvironment{proofof}{\noindent\sc Proof of}{
    \hspace*{\fill} $\Box$ \vspace{2ex} }

\newcommand{\Ind}[1]{{\boldsymbol{1}_{\textstyle\{#1\}}}}
\newcommand{\Indd}[1]{{\boldsymbol{1}_{\textstyle #1}}}

\def\eps{\varepsilon}

\def\Min(#1,#2){#1\wedge #2}
\def\Max(#1,#2){#1\vee #2}

\def\rueck{\noindent\hangafter=1 \hangindent=1.3em}

\begin{document}

\maketitle

\begin{abstract}
  We investigate the joint asymptotic behavior of so-called blocks
 estimator of the extremal index, that determines the mean length of
clusters of extremes, based on the exceedances over different
thresholds. Due to the large bias of these estimators, the resulting
estimates are usually very sensitive to the choice of the threshold
and thus difficult to
 interpret. We propose and examine a bias correction that asymptotically removes
 the leading bias term while the rate of convergence
 of the random error is preserved.
\end{abstract}


\section{Introduction}

When one analyzes a risk related to extreme values of a stationary
time series, then the clustering behavior of extremes can be as
least as important as the tail behavior of the marginal
distribution. For example, while a flood control basin may cope with
a single day of extreme rainfall, an extended period of heavy rain
will more likely lead to a flooding of the surrounding area.
Similarly, large negative returns on a stock index over several days
may sum up to an overall loss which is much worse than the most
extreme crash ever experienced on a single day.

Obviously, there is no single parameter which captures all facets of
serial dependence between extreme values, and in different
applications different features may be of interest. Recently, Drees
and Rootz\'{e}n (2010) introduced a very flexible class of empirical
processes that are capable of describing quite general aspects of
extremal dependence. In the present paper, it is demonstrated how
the asymptotic theory of these empirical processes can be used to
immensely improve the performance of well-known estimators of the
so-called extremal index, that is the reciprocal value of the
asymptotic mean cluster size.

More specifically, let a stationary time series $X_i, 1\le i\le n$,
with marginal distribution function (d.f.) $F$ be observed. We
assume that $F$ belongs to the maximum domain of attraction of some
extreme value d.f.\ $G_\gamma$, i.e., for an accompanying sequence
of independent and identically distributed (i.i.d.) random variables
(r.v.s) $\tilde X_i, 1\le i\le n$, with d.f.\ $F$ there exist
normalizing constants $a_n>0$ and $b_n\in\R$ such that
\begin{equation}  \label{eq:DOA}
P\Big\{ \frac{\max_{1\le i\le n} \tilde X_i -b_n}{a_n}\le x\Big\}
\;\longrightarrow\; G_\gamma(x),\quad x\in\R,
\end{equation}
as $n\to\infty$. It is well known  that (up to a scale and location
parameter) $G_\gamma$ must be of the form
$G_\gamma(x)=\exp\big(-(1+\gamma x)^{-1/\gamma}\big)$ for all $x$
such that $1+\gamma x>0$. Let
$$ u_n(x) := a_n x+b_n. $$
Moreover, assume the following mild mixing condition (a weakened
version of Leadbetter's condition $D$):

\hspace{0.5cm}\parbox{15cm}{There exist coefficients $\alpha_{n,l}$
and a sequence $l_n=o(n)$ such that $\alpha_{n,l_n}\to 0$ as
$n\to\infty$ and
$$ \big| P\big\{\max_{i\in I_1} X_i>u_n(x),\max_{i\in I_2}
X_i>u_n(x)\big\} - P\big\{\max_{i\in I_1} X_i>u_n(x)\}\cdot
P\big\{\max_{i\in I_2} X_i>u_n(x)\big\} \big|\le \alpha_{n,l}
$$
for all $x\in\R$ and all $I_1,I_2\subset \{1,\ldots,n\}$ such that
$\max I_1\le \min I_2-l$, $1\le l\le n-1$. }

Then there exists a constant $\theta\in[0,1]$, the so-called {\em
extremal index}, such that
\begin{equation}  \label{eq:maxdep}
  P\Big\{ \frac{\max_{1\le i\le n}  X_i -b_n}{a_n}\le x\Big\}
\;\longrightarrow\; G_\gamma^\theta(x),\quad x\in\R,
\end{equation}
provided that the left hand side converges (to an arbitrary limit)
for some $x\in\R$. In what follows, we will always rule out the
degenerate case $\theta=0$ which, in the limit, corresponds to
 clusters of extremes with infinite mean length.

If the extremal index $\theta$ is strictly positive, then usually it
may be interpreted as the reciprocal value of a limiting cluster
size. To see this, note that from \eqref{eq:DOA} and
\eqref{eq:maxdep} one may conclude
$$ \frac{P\{\max_{1\le i\le n} X_i>u_n(x)\}}{1-F^{n\theta}(u_n(x))}
\;\longrightarrow\; 1\quad \forall\, x\in\R,
$$
with the convention $0/0:=1$. Indeed, Hsing (1993, Theorem 3.1)
proved that under a stronger mixing condition this convergence holds
uniformly in $x$. If the following condition holds:

\hspace{0.5cm}\parbox{15cm}{There exist coefficients
$\tilde\alpha_{n,l}$ and a sequence $l_n=o(n)$ such that
$\tilde\alpha_{n,l_n}\to 0$ as $n\to\infty$ and
$$ \big| P\big(\max_{i\in I_2} X_i>u_n(x)\mid \max_{i\in I_1}
X_i>u_n(x)\big) - P\big\{\max_{i\in I_2} X_i>u_n(x)\big\} \big|\le
\tilde\alpha_{n,l}
$$
for all $x\in\R$ and all $I_1,I_2\subset \{1,\ldots,n\}$ such that
$\max I_1\le \min I_2-l$, $1\le l\le n-1$, }

then
\begin{equation}  \label{eq:unifconv}
  \sup_{u\in\R} \Big| \frac{P\{\max_{1\le i\le n}
  X_i>u\}}{1-F^{n\theta}(u)} -1\Big|
\;\longrightarrow\; 0.
\end{equation}
Now a Taylor expansion yields $1-F^{n\theta}(u)\sim n\theta \bar
F(u)=\theta E(C_n(u))$ uniformly for all $u\in[u_n,F^\leftarrow
(1))$ where $C_n(u):=\sum_{i=1}^n \Ind{X_i>u}$ denotes the total
number of exceedances over $u$, provided $u_n\to F^\leftarrow(1):=
\sup\{x\in\R\mid F(x)<1\}$ such that $\bar F(u_n) :=
1-F(u_n)=o(1/n)$. Hence, in view of \eqref{eq:unifconv}, it follows
\begin{equation}  \label{eq:recimean}
  \frac 1{E(C_n(u)\mid C_n(u)>0)} = \frac{P\{\max_{1\le i\le n}
  X_i>u\}}{n\bar F(u)} \;\longrightarrow\; \theta
\end{equation}
uniformly for all $u\in[u_n,F^\leftarrow(1))$.

Convergence \eqref{eq:recimean} suggests to estimate $\theta$ by
replacing the unknown probability and expectation on the left hand
side by empirical counterparts. Since we cannot estimate
$P\{\max_{1\le i\le n} X_i>u\}$ consistently if we observe merely
$n$ consecutive r.v.s $X_i$, $1\le i\le n$, we must first replace
$n$ with $r_n=o(n)$ in \eqref{eq:recimean} and adjust $u$
accordingly. Thus we split the sample into $m_n=\lfloor
n/r_n\rfloor$ blocks of length $r_n$ and estimate $\theta$ by
\begin{equation}  \label{eq:blockestdef}
  \hat\theta_n := \frac{\sum_{j=1}^{m_n}
  \Ind{\textstyle\max_{(j-1)r_n<i\le jr_n}
  X_i>u_n}}{\sum_{j=1}^{m_n} \sum_{i=(j-1)r_n+1}^{jr_n}
  \Ind{X_i>u_n}},
\end{equation}
for a sequence of thresholds $u_n$ satisfying $r_n\bar F(u_n)\to 0$,
but $n\bar F(u_n)\to\infty$.

This so-called {\em blocks estimator} of the extremal index has been
intensively studied in the literature. Hsing (1993) and Weissman and
Novak (1998) proved its consistency and asymptotic normality under
suitable mixing conditions. Variants of the blocks estimator were
also examined by Smith and Weissman (1994) and Robert et al.\
(2009). As alternatives to blocks estimators, so-called {\em runs
estimators} of $\theta$ have been proposed. While, in the numerator
of the right hand side of \eqref{eq:blockestdef}, the number of
clusters of extremes is defined as the number of blocks of length
$r_n$ which contain at least one exceedance, in the runs approach
two exceedances are considered to belong to different clusters if
they are separated by at least $\tilde r_n$ consecutive observations
that do not exceed $u_n$:
$$ \tilde\theta_n := \frac{\sum_{i=1}^{n-\tilde r_n}
  \Ind{X_i>u_n, X_j\le u_n \text{ for all } i+1\le j\le i+\tilde r_n
  }}{\sum_{i=1}^{n-\tilde r_n}  \Ind{X_i>u_n}}.
$$
The asymptotic behavior of this estimator was examined by Hsing
(1993), Smith and Weissman (1994) and Weissman and Novak (1998),
among others. Yet another approach was suggested by Ferro and Segers
(2003), who used interarrival times between exceedances to estimate
the extremal index.

In all these papers, the behavior of the estimators was analyzed for
a {\em fixed} sequence of thresholds. Below we will argue that the
analysis of the {\em joint} behavior of blocks estimators for
different thresholds does not only provide deeper insight, but that
it is the key to a remarkable reduction of the bias.

\begin{figure}[tb] \label{sampleplot}
\centerline{\includegraphics[height=50mm
]{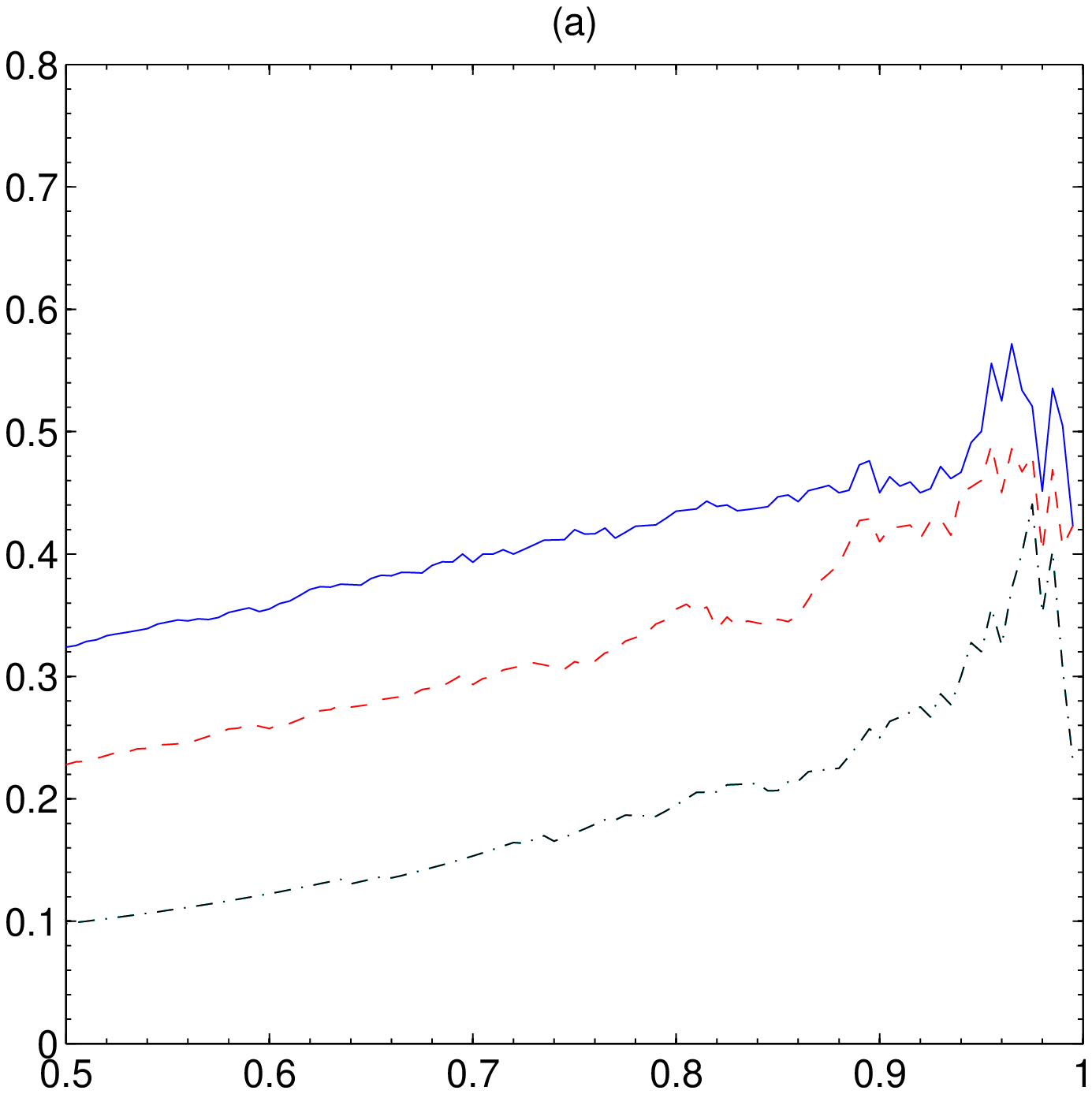}\quad
\includegraphics[height=50mm]{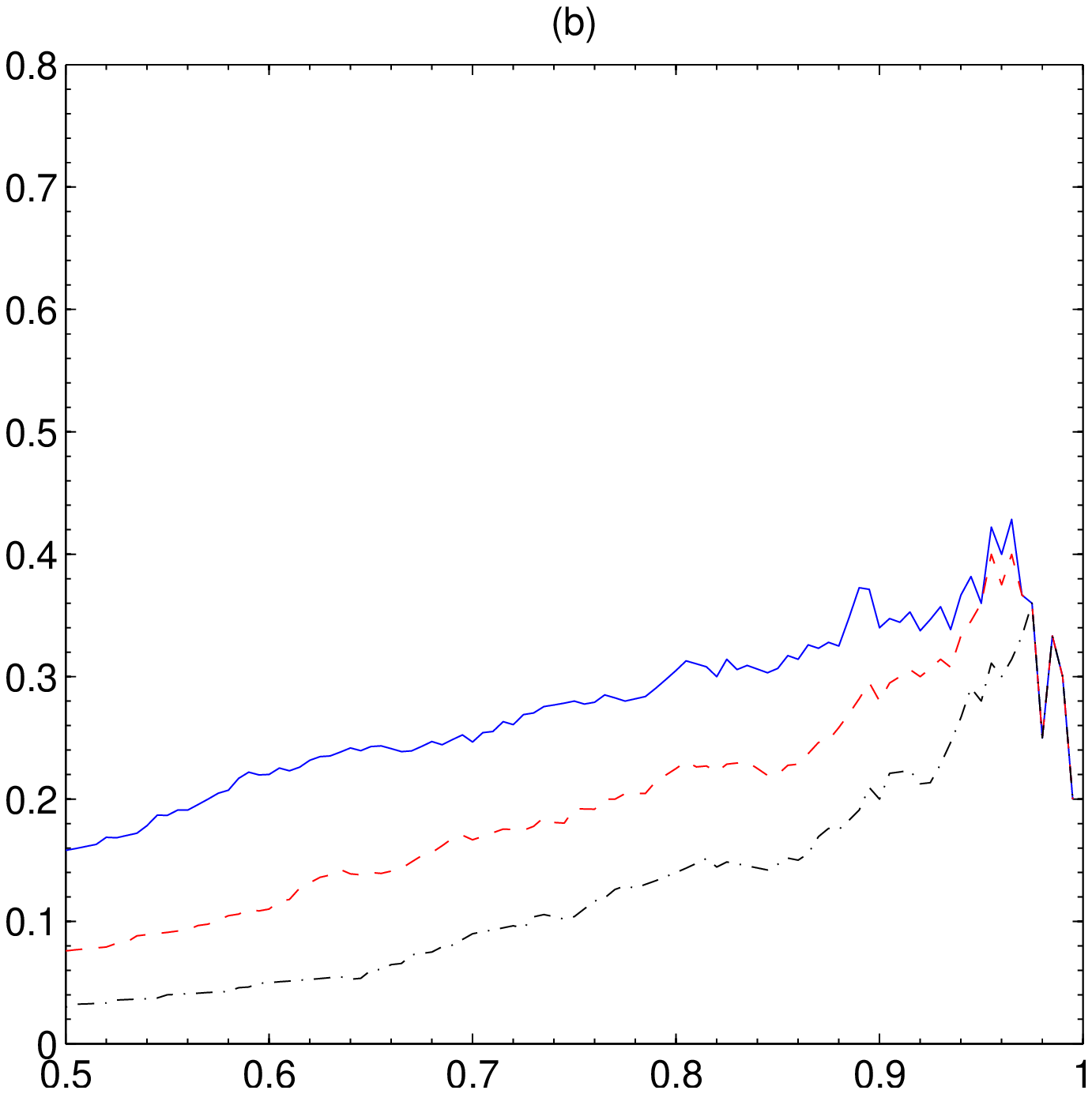}\quad
\includegraphics[height=50mm]{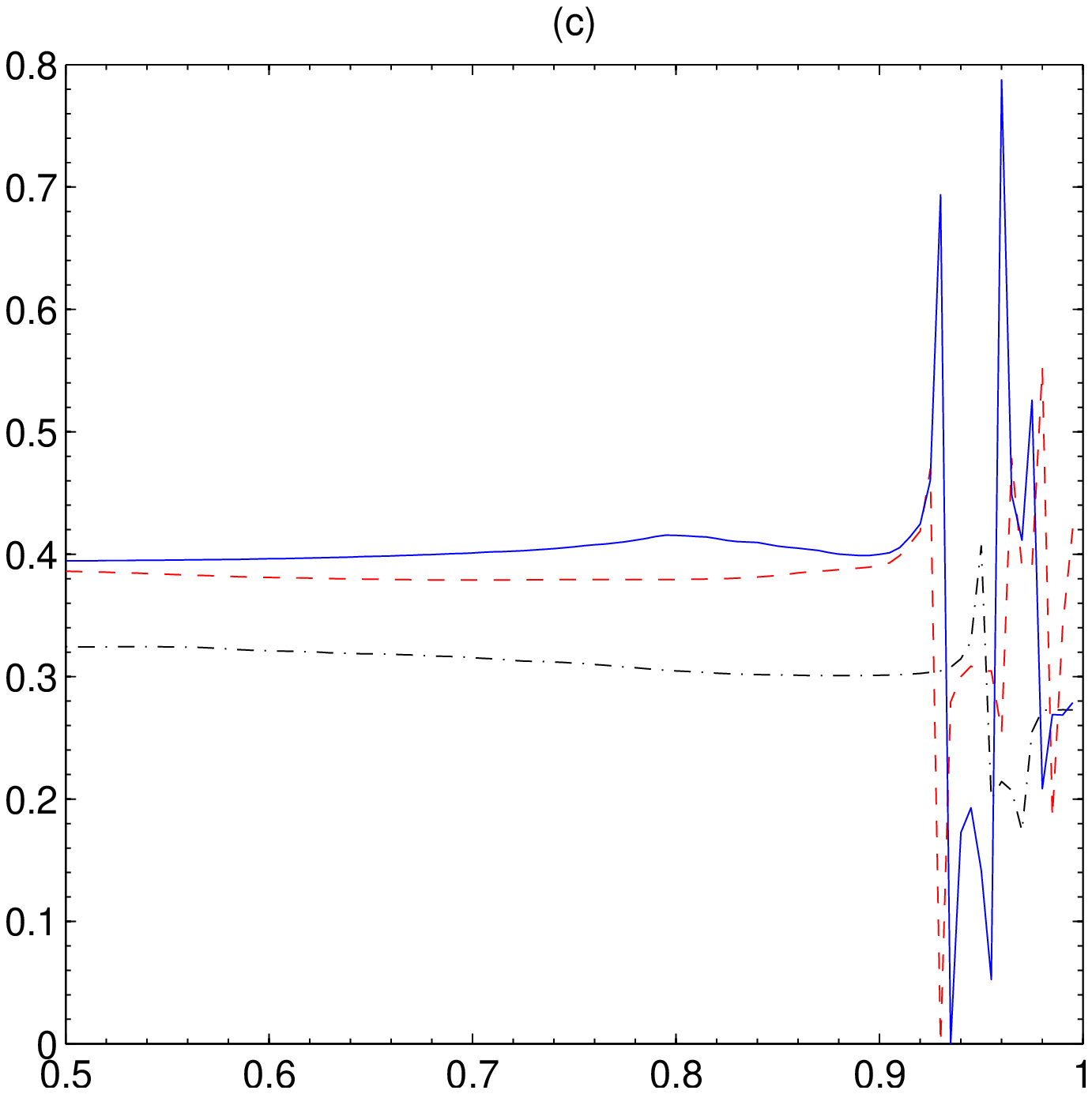}}\vspace{-0.1cm}
 \caption{Blocks estimator (left) with block lengths $r=5$ (blue solid line), $r=10$
 (red dashed) and $r=20$ (black dash-dotted), runs estimator (middle) with run lengths
 $\tilde r=2$ (blue solid), $\tilde r=5$
 (red dashed) and $\tilde r=10$ (black dash-dotted), and bias
 corrected blocks estimator (right) as functions of the standardized threshold for a AR(1)-times series with
 $\varphi=0.6$ and Cauchy innovations; the true extremal index equals
 $1-\varphi=0.4$.}
\end{figure}

Indeed, all the estimators mentioned above are plagued by serious
bias problems, which often renders inconclusive the analysis of the
strength of extremal dependence. As a typical example, consider the
following autoregressive time series of order 1 with Cauchy
innovations $\varepsilon_t$: $X_t=\varphi X_{t-1}+\eps_t$ with
$\varphi=0.6$. Figures \ref{sampleplot} (a) and (b) display blocks
and runs estimates of $\theta$ based on the exceedances over
$F_n^\leftarrow(u)=X_{n-\lceil nu\rceil+1:n}$ as a function of $u$
for several block lengths $r_n$, resp.\ run lengths $\tilde r_n$.
 (Here $F_n$ denotes the empirical d.f.\ and $X_{i:n}$
the $i$th smallest order statistic.)  The true value
$\theta=1-\varphi$ is indicated by the horizontal lines. The
estimates are almost monotone functions in $u$ and monotonically
increasing in the block lengths $r_n$, resp.\ run lengths $\tilde
r_n$. (The latter monotonicity holds by construction if $n$ is
divisible by $r_n$   resp.\ if the last $\tilde r_n$ observations do
not exceed the threshold.) Since there is no region where the
estimates remain stable, it is not obvious how to choose the
threshold appropriately.
Without an
objective procedure for choosing the threshold, it will thus be
difficult to justify any particular estimate for the extremal index.

In Section \ref{sect:biascorr} we suggest a method to combine blocks
estimators that are based on the exceedances over different
thresholds in a suitable way such that the leading bias term cancels
out for many well-known time series models. In Figure
\ref{sampleplot} (c) the resulting estimates based on exceedances
over $F_n^\leftarrow(u)$ are shown (again as a function of $u$) for
the same block lengths. Obviously, the estimates are not only almost
constant for a wide range of thresholds, but they also vary much
less with the block length than the original blocks estimator.

The remainder of the paper is organized as follows. In Section 2, we
establish a limit result for processes of blocks estimators indexed
by the threshold. To this end, we represent the blocks estimators as
functionals of a suitably defined empirical cluster process. Then
the joint asymptotic behavior of the blocks estimators easily
follows from a general limit theorem of such processes proved in
Drees and Rootz\'{e}n (2010). In the main Section \ref{sect:biascorr} we
first show that often the leading bias term of the blocks estimators
is a power function of the threshold. We then
 introduce a method to remove the leading bias
term of the blocks estimators in that case without deteriorating the
rate of convergence of the random error part. All proofs are
collected in Section \ref{sect:proofs}.

\section{Joint asymptotics of blocks estimators}

In this section we want to analyze the joint asymptotic behavior of
blocks estimators over a whole {\em continuum} of thresholds. Since
here we are interested in the extremal {\em dependence} (and not in
the marginal tails), the results should be invariant under strictly
increasing transformations of the observations. Hence it is natural
to parameterize the thresholds in terms of the marginal quantile
function $F^\leftarrow$, that is to consider
$$ \hat\theta_{n,t}^* := \frac{ \sum_{j=1}^{m_n} \Ind{
\max_{(j-1)r_n<i\le jr_n}
X_i>F^\leftarrow(1-v_nt)}}{\sum_{j=1}^{m_n}
\sum_{i=(j-1)r_n+1}^{jr_n}\Ind{X_i>F^\leftarrow(1-v_nt)}}, \quad
0<t\le 1.
$$

For later applications, though, it is more convenient to examine a
version where the unknown quantile function is replaced with an
empirical analog:
$$ \hat\theta_{n,t} := \frac{ \sum_{j=1}^{m_n} \Ind{
\max_{(j-1)r_n<i\le jr_n} X_i>X_{n-\lceil
nv_nt\rceil:n}}}{\sum_{j=1}^{m_n}
\sum_{i=(j-1)r_n+1}^{jr_n}\Ind{X_i>X_{n-\lceil nv_nt\rceil:n}}},
\quad 0<t\le 1.
$$
If there are no ties among the largest $\lceil nv_n\rceil$
observations and none of them are among the last $n-m_nr_n$
observations, then $\hat\theta_{n,t}$ can be rewritten as
$$ \hat\theta_{n,t} = \frac 1{\lceil nv_nt\rceil} \sum_{j=1}^{m_n} \Ind{
\max_{(j-1)r_n<i\le jr_n} X_i>X_{n-\lceil nv_nt\rceil:n}}.
$$
In particular, this representation holds with probability tending to
1 if we assume that $F$ is continuous on some neighborhood of
$F^\leftarrow(1)$ and $r_nv_n\to 0$, which we will do throughout the
remainder of the paper.

For sufficiently large $n$, we then have
$$ \hat\theta_{n,t}^* = \frac{ \sum_{j=1}^{m_n} \Ind{
\max_{(j-1)r_n<i\le jr_n} U_i>1-v_nt}}{\sum_{j=1}^{m_n}
\sum_{i=(j-1)r_n+1}^{jr_n}\Ind{U_i>1-v_nt}}
$$
where the random variables $U_i=F(X_i)$, $1\le i\le n$, have a
distribution which equals the uniform distribution in a neighborhood
of 1. Thus this blocks estimator can be expressed in terms of
certain empirical processes of cluster functionals that have been
introduced and analyzed by Drees and Rootz\'{e}n (2010). To this end,
define standardized excesses
$$ U_{n,i} := \frac {(U_i-(1-v_n))^+}{v_n} = \frac
{(U_i-(1-v_n))\vee 0}{v_n}, \quad 1\le i\le n,
$$
blocks thereof
$$ Y_{n,j} := (U_{n,i})_{(j-1)r_n<i\le j r_n}, \quad 1\le j\le m_n,
$$
and functionals on $\R_\cup:=\bigcup_{l\in\N} \R^l$ by
\begin{eqnarray*}
  f_t(x_1,\ldots,x_l) & := & \Ind{\max_{1\le i\le l}
  x_i>1-t}\\
  g_t(x_1,\ldots,x_l) & := & \sum_{i=1}^l \Ind{
  x_i>1-t}.
\end{eqnarray*}
Then
\begin{equation}
 \hat\theta_{n,t}^*  =  \frac{m_n^{-1} \sum_{j=1}^{m_n}
 f_t(Y_{n,j})}{m_n^{-1} \sum_{j=1}^{m_n}
 g_t(Y_{n,j})}  =  \frac{E(f_t(Y_{n,1}))+(nv_n)^{1/2} m_n^{-1} Z_n(f_t)}{E(g_t(Y_{n,1}))+(nv_n)^{1/2} m_n^{-1}
  Z_n(g_t)},
    \label{eq:empprocrep}
\end{equation}
where for a generic functional $h:\R_\cup\to\R$ we define
$$ Z_n(h) := \frac 1{\sqrt{nv_n}} \sum_{j=1}^{m_n}\big( h(Y_{n,j})-E
h(Y_{n,j})\big).
$$
Under suitable conditions on the time series and the family
$\mathcal{H}$ of functionals $h$, Drees and Rootz\'{e}n (2010) proved
convergence of the empirical processes $(Z_n(h))_{h\in\mathcal{H}}$
to a centered Gaussian process with continuous sample paths.

Here we recall conditions that ensure the convergence of the
processes $\big(Z_n(f_t),Z_n(g_t)\big)_{0\le t\le 1}$. Note that
$(Z_n(g_t))_{0\le t\le 1}$ is the usual tail empirical process,
whose asymptotic behavior has been investigated by Rootz\'{e}n (1995,
2009) and Drees (2000).

\vspace{1ex} {\bf(C1)} \parbox[t]{15cm}{The $\beta$-mixing
coefficients
$$ \beta_{n,k} := \sup_{1\le l\le n-k-1} E \Big(
\sup_{B\in\BB_{n,l+k+1}^n} \big| P(B|\BB_{n,1}^l)-P(B)\big|\Big)
$$
of the vector of excesses $(X_k-F^\leftarrow(1-v_n(1-\eps)))^+_{1\le
k\le n}$ satisfy $\beta_{n,l_n} n/r_n\to 0$ for some sequence
$l_n=o(r_n)$. Here $\BB_{n,i}^j$ denotes the $\sigma$-field
generated by $(X_k-F^\leftarrow(1-v_n(1-\eps)))^+_{i\le k\le j}$ for
some $\eps>0$.}

 \vspace{1ex} {\bf(C2)} \parbox[t]{15cm}{$r_n\to\infty$, $r_nv_n\to
 0$, $nv_n\to\infty$.}

 \vspace{1ex} {\bf(C3.1)} \parbox[t]{15cm}{For some $\eps>0$
 \begin{eqnarray*}
    \lefteqn{\frac 1{r_nv_n} Cov\Big( \sum_{i=1}^{r_n} \Ind{X_i>F^\leftarrow(1-v_n(1-s))},\sum_{i=1}^{r_n}
 \Ind{X_i>F^\leftarrow(1-v_n(1-t))}\Big)}\\
  & \to & c_g(s,t) \qquad \forall\, -\eps\le
 s,t\le 1. \hspace*{7cm}
 \end{eqnarray*}}

 \vspace{1ex} {\bf(C3.2)} \parbox[t]{15cm}{For some $\eps>0$
 \begin{eqnarray*}
    \lefteqn{\frac 1{r_nv_n} Cov\Big(\Ind{\max_{1\le i\le r_n}X_i>F^\leftarrow(1-v_n(1-s))},\sum_{i=1}^{r_n}
 \Ind{X_i>F^\leftarrow(1-v_n(1-t))}\Big)}\\
  & \to &c_{fg}(s,t) \qquad \forall\, -\eps\le
 s,t\le 1. \hspace*{8cm}
 \end{eqnarray*}}

 \vspace{1ex} {\bf(C4)} \parbox[t]{15cm}{There exists a bounded
 function $h:(0,1]\to\R$ such that $\lim_{t\to 0}h(t)=0$ and for
 sufficiently large $n$
 $$ \frac 1{r_nv_n} E\Big( \sum_{i=1}^{r_n} \Ind{F^\leftarrow(1-v_n(1-s))<X_i\le F^\leftarrow(1-v_n(1-t))}\Big)^2
 \le h(t-s)
  \quad \forall\, -\eps\le s<t\le 1.
 $$}

 \begin{theorem} \label{theo:empproc}
   \begin{enumerate}
     \item Under the conditions (C1) and (C2), $(Z_n(f_t))_{0\le
     t\le 1}$ converge weakly to $Z_f:=(\sqrt{\theta} B_t)_{0\le
     t\le 1}$ with $B$ denoting a standard Brownian motion.
     \item If the conditions (C1), (C2), (C3.1) and (C4) are met and
     $r_n=o(\sqrt{nv_n})$, then $(Z_n(g_t))_{0\le t\le 1}$ converge
     to a centered Gaussian process $(Z(g_t))_{0\le t\le 1}$ with
     covariance function $c_g$.
     \item If the     conditions (C1)--(C4) are satisfied and
     $r_n=o(\sqrt{nv_n})$, then $(Z_n(f_t),Z_n(g_t))_{0\le t\le
     1}$ converge weakly to $(Z_f(t),Z_g(t))_{0\le t\le
     1}$ with
     \begin{eqnarray*}
       Cov(Z_f(s),Z_f(t)) & = & \theta(s\wedge t),\\
       Cov(Z_g(s),Z_g(t)) & = & c_g(s,t),\\
       Cov(Z_f(s),Z_g(t)) & = & c_{fg}(s,t),\quad 0\le s,t\le 1.
     \end{eqnarray*}
   \end{enumerate}
 \end{theorem}
 \begin{remark}
   The covariance conditions (C3.1) and (C3.2) are fulfilled if all
   finite dimensional marginal distributions $(X_1,\ldots,X_k)$
   belong to the domain of attraction of some multivariate extreme
   value distribution, $\lim_{n\to\infty} \limsup_{m\to\infty} \beta_{n,m}=0$,
    and the following condition holds:

  \vspace{1ex} {\bf(C5)} \parbox[t]{15cm}{For some $\delta>0$
 $$   E\Big( \sum_{i=1}^{r_n} \Indd{(0,1]}(U_{n,i})\Big)^{2+\delta}
 =O(r_nv_n).
 $$}

   In this case, Segers (2003) has shown that the conditional
   distributions $P^{(U_{n,i})_{1\le i\le k}|U_{n,1}\ne 0}$ of $(U_{n,i})_{1\le i\le
   k}$ given that the first observation exceeds the threshold
   converge weakly to the distribution of $(W_i)_{1\le i\le
   k}=(V_i\vee 0)_{1\le i\le k}$, where $(V_i)_{1\le i\le k}$ is
   the so-called tail sequence pertaining to the
   time series $U_i$, $i\in\N$. The limiting covariance functions
   $c_g$ and $c_{fg}$ are then given by
   \begin{eqnarray}
     c_g(s,t) & = & s\wedge t + \sum_{k=2}^\infty\big( P\{W_1>1-s,
     W_k>1-t\}+P\{W_1>1-t,     W_k>1-s\} \big),   \label{eq:cg}\\
     c_{fg}(s,t) & = & \left\{ \begin{array}{l@{\quad}l}
       P\{W_1>1-t,\max_{j\ge 1}W_j>1-s\}\\
        \hspace*{1em}+\sum_{k=2}^\infty P\{W_1>1-s,W_k>1-t,\max_{j\ge 2}W_j\le
       1-s\}, & s<t,\\
       t & s\ge t.
       \end{array} \right.  \label{eq:cfg}
   \end{eqnarray}
   { }
 \end{remark}

Using the joint convergence of $Z_n(f_t)$ and $Z_n(g_t)$ and the
representation \eqref{eq:empprocrep}, one can easily derive a limit
theorem for the processes $(\hat\theta_{n,t}^*)_{0< t\le 1}$ of
blocks estimators.
\begin{corollary} \label{cor:blockasymp}
  Under the conditions of Theorem \ref{theo:empproc} (iii)
  $$ \big( \sqrt{nv_n}
  t(\hat\theta_{n,t}^*-\theta_{n,t})\big)_{0<t\le 1} \;\to\;
  Z:=Z_f-\theta Z_g \quad\text{weakly as } n\to\infty
  $$
  with
  $$ \theta_{n,t} := \frac{E(f_t(Y_{n,1}))}{E(g_t(Y_{n,1}))} =
\frac{P\{\max_{1\le i\le r_n} X_i>F^\leftarrow (1-v_nt)\}}{r_nv_nt}.
$$
 The limit process $Z$ is Gaussian with $E(Z(t))=0$ and
 \begin{equation} \label{eq:covdef}
  Cov(Z(s),Z(t)) = \theta\big(s\wedge
  t-c_{fg}(s,t)-c_{fg}(t,s)\big)+\theta^2 c_g(s,t) =: c(s,t).
 \end{equation}
\end{corollary}
Note that the centering constant $\theta_{n,t}$, which is the
leading term in the representation \eqref{eq:empprocrep}, converges
to $\theta$ uniformly for all $t\in(0,1]$ by Hsing's (1993) result
\eqref{eq:recimean}. However, the convergence can be rather slow
leading to a large bias of the blocks estimator as observed in
Figure \ref{sampleplot}.

In the next section we will see how to combine all blocks estimators
$\hat\theta_{n,t}^*$ non-linearly such that the resulting estimator
has a much smaller bias. As the threshold $F^\leftarrow(1-v_nt)$ is
unknown, for any given threshold $u_n$ in the definition
\eqref{eq:blockestdef} it is not known for which index $t$ one has
$\hat\theta_n =\hat\theta_{n,t}^*$. Hence, we first need an analog
to Corollary \ref{cor:blockasymp} for the estimator
$\hat\theta_{n,t}$ with random threshold $X_{n-\lceil
nv_nt\rceil:n}$.

To this end, we analyze the difference between the deterministic
threshold $1-v_nt$ (after standardization of the marginals) and its
random counterpart $1-U_{n-\lceil nv_nt\rceil:n}$. It has been shown
in Drees (2000), proof of Corollary 3.1, that
$\sqrt{nv_n}\big((1-U_{n-\lceil nv_nt\rceil:n})/v_n-t\big)_{0\le
t\le 1}$ converges to a Gaussian process if $(Z_n(g_t))_{-\eps\le
t\le 1}$ converges to a Gaussian process. Note that the latter
convergence follows from an analog to Theorem \ref{theo:empproc}
(ii), because the conditions (C3.1) and (C4) have been formulated
for $s,t\in[-\eps,1]$ (while for Theorem \ref{theo:empproc} (ii) to
hold it suffices to require the conditions for $s,t\in[0,1]$). This
suffices to establish a limit theorem for $\hat\theta_{n,t}$. It
turns out that under a suitable continuity condition on
$\theta_{n,t}$, the blocks estimator with estimated threshold has
the same asymptotic behavior as $\hat\theta_{n,t}^*$.
\begin{corollary} \label{cor:blockasympest}
  Suppose the conditions of Theorem \ref{theo:empproc} (iii) are
  met. Then
  \begin{equation} \label{eq:blockpreasympest}
   \big( \sqrt{nv_n}
   t(\hat\theta_{n,t}-\theta_{n,(1-U_{n-\lceil
nv_nt\rceil:n})/v_n})\big)_{0\le t\le 1} \;\to\;
  Z \quad\text{weakly as } n\to\infty.
  \end{equation}

  If, in addition, to each $t_0\in (0,1)$ and each $M_1>0$ there exists $M_2>0$ such that
  \begin{equation} \label{eq:thetantcont}
    \sup_{s,t\ge t_0, |s-t|\le M_1(nv_n)^{-1/2}} \Big|
    \frac{\theta_{n,s}-\theta}{\theta_{n,t}-\theta}-1\Big| \le
    M_2(nv_n)^{-1/2},
  \end{equation}
  then
  \begin{equation}  \label{eq:blockasympest}
   \big( \sqrt{nv_n}
    t(\hat\theta_{n,t}-\theta_{n,t})\big)_{0\le t\le 1} \;\to\;
  Z \quad\text{weakly as } n\to\infty.
  \end{equation}
\end{corollary}

\section{Bias correction}
\label{sect:biascorr}

As in Figure \ref{sampleplot}, the blocks estimator
$\hat\theta_{n,t}$ often exhibits a clear trend, that is caused by
its bias, when it is plotted versus $t$. In this section we show how
to combine blocks estimators for different thresholds such that the
leading bias term vanishes while the order of magnitude of the
random error is preserved. To this end, we make structural
assumptions on the form of the bias $\theta_{n,t}-\theta$ as a
function of $t$. The following examples demonstrate that in
 time series models discussed in the literature the leading bias term often
equals a power of $t$ with positive exponent.

\begin{example} \label{ex:Weissman} \rm
  Let $Z_i$, $i\in\N$, be iid r.v.s with d.f.\ $F$, and let $\xi_i$,
  $i\in\N$, denote a series of iid Bernoulli rvs, independent of
  $(Z_i)_{i\in\N}$, with
  $P\{\xi_i=0\}=\psi=1-P\{\xi_i=1\}$. Weissman and Novak (1998, p.\ 285) proved that
  then the time series $X_0:=Z_0$, $X_t:=\xi_t
  Z_t+(1-\xi_t)X_{t-1}$, $t\in\N$, is stationary with marginal d.f.\
  $F$ and extremal index $\theta=1-\psi$. Moreover, if $F$ is
  eventually continuous, then for all $t_0>0$
  $$ \theta_{n,t} = \frac{1-(1-v_nt)(1-\theta v_nt)^{r_n-1}}{r_nv_n
  t} = \theta -\frac{\theta^2}2 r_nv_n t+\frac{1-\theta}{r_n} +
  O(v_n+r_n^2v_n^2)
  $$
  uniformly for $t\in[t_0,1]$. If $r_n^2v_n\to\infty$, then the linear function $-\theta^2 r_nv_n
  t/2$ is the leading bias term.
\end{example}

\begin{example} \label{ex:movingmaxima} \rm
  Consider a finite order moving maxima time series
  $$ X_t= \max_{0\le j\le q} (\psi_jZ_{t-j}) $$
  with non-negative coefficients $\psi_j\ge 0$. W.l.o.g.\ we may and
  will assume that $ \max_{0\le j\le q} \psi_j=1$.   Further assume
  that the innovations $Z_t$ are iid with heavy tailed d.f.\ $F_Z$
  satisfying
  $$ \bar F_Z(z) := 1-F_Z(z)= c_1z^{-\beta_1}\big(1+c_2
  z^{-\beta_2} + o(z^{-\beta_2})\big)
  $$
  for some $\beta_1,\beta_2,c_1>0$ and $c_2\ne 0$.

  If $\beta_2<\beta_1$, then
  \begin{eqnarray*}
    F(x) & := & P\{X_t\le x\}\\
     & = & P\{Z_{t-j}\le x/\psi_j\; \forall\, 0\le j\le q\} \\
     & = &
     \prod_{j=0}^q\Big(1-c_1(x/\psi_j)^{-\beta_1}\big(1+c_2(x/\psi_j)^{-\beta_2}+o(x^{-\beta_2})\big)\Big)\\
     & = & 1- c_1 \sum_{j=0}^q \psi_j^{\beta_1}x^{-\beta_1} - c_1c_2
     \sum_{j=0}^q \psi_j^{\beta_1+\beta_2} x^{-(\beta_1+\beta_2)} +
     o\big(x^{-(\beta_1+\beta_2)}\big)
  \end{eqnarray*}
  as $x\to\infty$, and thus for all fixed $A>0$
  \begin{eqnarray*}
    \bar F_Z(x/A) & = & \frac{A^{\beta_1}}{\sum_{j=0}^q
  \psi_j^{\beta_1}} \bar F(x) + \frac{c_2}{c_1^{\beta_2/\beta_1}}
  \frac{A^{\beta_1+\beta_2}}{\big(\sum_{j=0}^q
  \psi_j^{\beta_1}\big)^{1+\beta_2/\beta_1}} \Big( 1- \frac{\sum_{j=0}^q
  \psi_j^{\beta_1+\beta_2}}{\sum_{j=0}^q
  \psi_j^{\beta_1}A^{\beta_2}}\Big) (\bar F(x))^{1+\beta_2/\beta_1}\\
  & & +
  o(x^{(-\beta_1+\beta_2)}).
  \end{eqnarray*}

  To determine $\theta_{n,t}$, check that with
  $$ d := \frac{c_2}{c_1^{\beta_2/\beta_1}}
  \frac1{\big(\sum_{j=0}^q
  \psi_j^{\beta_1}\big)^{1+\beta_2/\beta_1}} \Big( 1- \frac{\sum_{j=0}^q
  \psi_j^{\beta_1+\beta_2}}{\sum_{j=0}^q
  \psi_j^{\beta_1}}\Big)
  $$
  it follows that
  \begin{eqnarray*}
    \lefteqn{P\big\{\max_{1\le t\le r_n} X_t\le F^\leftarrow(1-v_n
    t)\big\}}\\
    & = & P\Big\{ Z_{t-j} \le \frac{F^\leftarrow(1-v_nt)}{\psi_j} \;
    \forall\, 1\le t\le r_n, 0\le j\le q\Big\} \\
    & = & P\Big\{ Z_m \le \frac{F^\leftarrow(1-v_nt)}{\max_{0\vee (1-m)\le
    j\le q \wedge (r_n-m)} \psi_j}\; \forall\, 1-q\le m\le r_n\Big\}
    \\
    & = & \prod_{m=1-q}^0
    F_Z\Big(\frac{F^\leftarrow(1-v_nt)}{\max_{1-m\le j\le q}
    \psi_j}\Big) \cdot \prod_{m=1}^{r_n-q}
    F_Z\Big(\frac{F^\leftarrow(1-v_nt)}{\max_{0\le j\le q}
    \psi_j}\Big) \cdot \prod_{m=r_n-q+1}^{r_n}
    F_Z\Big(\frac{F^\leftarrow(1-v_nt)}{\max_{0\le j\le r_n-m}
    \psi_j}\Big)\\
    & = & \prod_{m=1-q}^0 (1+O(v_n))\cdot \bigg( 1- \frac
    1{\sum_{j=0}^q
  \psi_j^{\beta_1}} v_n t - d (v_nt)^{1+\beta_2/\beta_1}+
  o(v_n^{1+\beta_2/\beta_1})\Big)^{r_n-q}\\
  & &  \cdot  \prod_{m=r_n-q+1}^{r_n}(1+O(v_n))\\
    & = & 1- \frac    1{\sum_{j=0}^q
  \psi_j^{\beta_1}} r_n v_n t - dr_n(v_nt)^{1+\beta_2/\beta_1}+
  O(v_n+r_n^2v_n^2)+
  o(r_nv_n^{1+\beta_2/\beta_1}).
  \end{eqnarray*}
  Hence,
  if $r_nv_n^{\beta_2/\beta_1}\to\infty$ but
  $r_nv_n^{1-\beta_2/\beta_1}\to 0$ (which implies
  $\beta_2<\beta_1/2$), then for all $t_0>0$
  $$ \theta_{n,t} = \frac{1-P\big\{\max_{1\le t\le r_n} X_t\le F^\leftarrow(1-v_n
    t)\big\}}{r_nv_n t} = \frac 1{\sum_{j=0}^q  \psi_j^{\beta_1}} + d(v_nt)^{\beta_2/\beta_1} +
  o(v_n^{\beta_2/\beta_1})
  $$
  uniformly for $t\in [t_0,1]$. Here the the constant $d$ is strictly
  negative if $\psi_j\in (0,1)$ for some $j\in\{0,\ldots,q\}$.
  Hence, in this case,  $\theta:= 1/\sum_{j=0}^q  \psi_j^{\beta_1}$ is the extremal index and
   the leading term  of the bias
  $\theta_{n,t}-\theta$ is a multiple of $t^{\beta_2/\beta_1}$.

  Now we investigate the general case, i.e.\ we do not assume that
  $\beta_2<\beta_1$. By similar calculations as above, we obtain
  that
  \begin{eqnarray*}
   \lefteqn{\bar F(x) = 1+c_1 \sum_{j=0}^q  \psi_j^{\beta_1}
   x^{-\beta_1} + O(x^{-(\beta_1+\beta_2)}+x^{-2\beta_1})}\\
   & \Longrightarrow & \bar F_Z(x/A) = \frac{A^{\beta_1}}{\sum_{j=0}^q
  \psi_j^{\beta_1}} \bar F(x) +
  O(x^{-(\beta_1+\beta_2)}+x^{-2\beta_1}).
  \end{eqnarray*}
  Therefore
  \begin{eqnarray*}
    \lefteqn{P\big\{\max_{1\le t\le r_n} X_t\le F^\leftarrow(1-v_n
    t)\big\}}\\
    & = & (1+O(v_n))^{2q}\cdot\Big( 1-\frac{v_nt}{\sum_{j=0}^q
    \psi_j^{\beta_1}} +
    O(v_n^{1+\beta_2/\beta_1}+v_n^2)\Big)^{r_n-q}\\
    & = & 1- \frac{r_nv_nt}{\sum_{j=0}^q  \psi_j^{\beta_1}} + \frac
    12\bigg(\frac{r_nv_nt}{\sum_{j=0}^q  \psi_j^{\beta_1}}\bigg)^2 +
    O\big(v_n+(r_nv_n)^3+r_n(v_n^{1+\beta_2/\beta_1}+v_n^2)\big),
  \end{eqnarray*}
  which in turn implies
  $$ \theta_{n,t} = \theta-\frac{\theta^2}2r_nv_nt+o(r_nv_n)
  $$
  if $r_nv_n^{\max(1/2,1-\beta_2/\beta_1)}\to \infty$. Hence, in
  this case the leading bias term is a linear function of $t$.
\end{example}
\begin{remark}
  Theorem 4.1 of Hsing (1993) suggests that indeed for $m$-dependent
  time series with $m$-dimensional regularly varying marginal
  distributions the leading bias term  usually is a linear
  function of $t$ if $r_n\to\infty$ sufficiently fast.
\end{remark}

We propose the following estimator of the extremal index with
reduced bias:
\begin{equation} \label{eq:redbiasest}
   \hat \theta_{n,\mu} := \frac{\displaystyle \int_{(0,1]^2} \hat\theta_{n,s}
   \hat\theta_{n,t}\, \mu(ds,dt)}{\displaystyle \int_{(0,1]^2} \hat\theta_{n,s}+
   \hat\theta_{n,t}\, \mu(ds,dt)},
\end{equation}
where $\mu$ is some finite signed measure on $(0,1]^2$ satisfying
the following conditions:

\vspace{1ex} {\bf(M1)} \parbox[t]{15cm}{The signed measure $\mu^\pi$
induced by the product map $\pi:(0,1]^2\to (0,1]$, $\pi(s,t)=st$,
vanishes, i.e.\ $\mu\{\pi\in B\}=0$ for all $B\in\B((0,1])$.}

\vspace{1ex} {\bf(M2)} \parbox[t]{15cm}{$\int
s^\delta+t^\delta\,\mu(ds,dt)\ne 0$ for all $\delta>0$.}

\vspace{1ex} {\bf(M3)} \parbox[t]{15cm}{The total variation measure
$|\mu|$ pertaining to $\mu$ satisfies $\int_{(0,1]^2}
(st)^{-1}\,|\mu|(ds,dt)<\infty$.}

\vspace{1ex}

\begin{example}  \label{ex:mu} \rm
  \begin{enumerate}
    \item
  Let $F,G$ be d.f.s of probability measures $Q_F$ and $Q_G$ on $(0,1]$ such
  that\newline
   $\int_{(0,1]} t^{-1} \, Q_F(dt)<\infty$, $\int_{(0,1]} t^{-1} \, Q_G(dt)<\infty$ and
   $\int_{(0,1]} t^\delta \, Q_F(dt)\ne \int_{(0,1]} t^\delta \,
   Q_G(dt)$ for all $\delta>0$. (The latter condition is, for instance, fulfilled
   if $Q_F$ equals the distribution $Q_G^{T_b}$ of the map $T_b(x):=x/b$ under
   $Q_G$ for some $b>1$.) Then the signed measure
   $\mu=Q_F^{T_a}\otimes Q_G-Q_F\otimes Q_G^{T_a}$ for some $a>1$ (i.e.,
   $\mu\big((0,x]\times(0,y]\big)=F(ax)G(y)-F(x)G(ay)$ for all
   $x,y\in(0,1]$) satisfies the conditions (M1)--(M3):
  \begin{eqnarray*}
    \mu\{(s,t)\mid st\le u\} & = & \int_{(0,1]} G(u/s)\, Q_F^{T_a}(ds)-\int_{(0,1]} G(au/s)\, Q_F(ds)=0,\\
    \int_{(0,1]^2} s^\delta+t^\delta\, \mu(ds,dt)
     & = &  \int_{(0,1]} s^\delta\, Q_F^{T_a}(ds) + \int_{(0,1]} t^\delta\, Q_G(dt)
     -\int_{(0,1]} s^\delta\, Q_F(ds) - \int_{(0,1]} t^\delta\, Q_G^{T_a}(dt)\\
    & = & (a^{-\delta}-1) \Big(\int_{(0,1]} s^\delta\, Q_F(ds)-\int_{(0,1]} t^\delta\, Q_G(dt)\Big)\\
    & \ne & 0,\\
     \int_{(0,1]^2} (st)^{-1}\, |\mu|(ds,dt) & = & \int_{(0,1]} s^{-1}\,
     Q_F^{T_a}(ds) \int_{(0,1]} t^{-1}\, Q_G(dt)+ \int_{(0,1]} s^{-1}\,
     Q_F(ds) \int_{(0,1]} t^{-1}\, Q_G^{T_a}(dt)\\
     & = & (1+a)\int_{(0,1]} s^{-1}\,
     Q_F(ds)\cdot \int_{(0,1]} t^{-1}\, Q_G(dt)<\infty.
  \end{eqnarray*}

  \item The above example is a special case of the following more
  general construction. Let $T:(0,1]^2\to D:=\{(u,v)\mid 0<u\le
  v\le 1\}$, $T(x,y):=(xy,y)$, and let $T^{-1}:D\to (0,1]^2$, $T^{-1}(u,v)=(u/v,v)$ denote
  its inverse. Choose some measure $\nu$ on
  $(0,1]$ satisfying $\int_{(0,1]} s^{-1}\,\nu(ds)<\infty$, and Markov kernels $K_1$ and $K_2$ from $(0,1]$ to
  $(0,1]$ such that $K_i(u,[u,1])=1$. Then the signed measure
  $\mu:=(\nu\otimes K_1)^{T^{-1}}-(\nu\otimes
  K_2)^{T^{-1}}$ meets the conditions (M1) and (M3), because  $\pi=T\circ pr_1$ with $pr_1$ denoting the
  projection on the first coordinate, and thus $\mu^\pi=\Big(\big((\nu\otimes
  K_1)^{T^{-1}}\big)^T\Big)^{pr_1}-\Big(\big((\nu\otimes
  K_2)^{T^{-1}}\big)^T\Big)^{pr_1} = \nu-\nu=0$ and $\int_{(0,1]^2}
  (st)^{-1} \, (\nu\otimes K_i)^{T^{-1}}(ds,dt) = \int_{(0,1]^2} u^{-1}\,
  (\nu\otimes K_i) (du,dv)<\infty$.
 \end{enumerate}

\end{example}

Our main result shows that the bias of $\hat\theta_{n,\mu}$ (and
hence its estimation error) is of smaller order than the bias of
$\theta_{n,t}$ if the bias dominates the random error and its
leading term is a power function.
\begin{theorem}  \label{theo:main}
  Suppose that conclusion \eqref{eq:blockasympest} of Corollary
  \ref{cor:blockasympest} holds and that
  \begin{equation}  \label{eq:biascond}
     \theta_{n,t}=\theta_n+c_nt^\delta + R_n(t) \quad \forall\,
     t\in (0,1]
  \end{equation}
  for  some $\delta>0$ with $d_n:=\sup_{0< t\le 1}t|R_n(t)|=o(c_n)$ and $(nv_n)^{-1/2}=o(c_n)$.  If the conditions
  (M1)--(M3) are fulfilled, then
  \begin{eqnarray*}
    \hat\theta_{n,\mu} & =^d & \theta_n+(nv_n)^{-1/2}
  \frac{\displaystyle \int_{(0,1]^2} s^\delta t^{-1} Z(t)+t^\delta s^{-1} Z(s)\,
  \mu(ds,dt)}{\displaystyle \int_{(0,1]^2} s^\delta +t^\delta\,
  \mu(ds,dt)}\\
  & &
  + \frac{\displaystyle \int_{(0,1]^2} s^\delta R_n(t)+t^\delta R_n(s)\,
  \mu(ds,dt)}{\displaystyle \int_{(0,1]^2} s^\delta +t^\delta\,
  \mu(ds,dt)} + o_P((nv_n)^{-1/2}+d_n).
  \end{eqnarray*}
  In particular, if $d_n=o((nv_n)^{-1/2})$,
  then
  \begin{equation}  \label{eq:main}
    \sqrt{nv_n}(\hat\theta_{n,\mu} -\theta_n) \;\longrightarrow\;
    \frac{\displaystyle \int_{(0,1]^2} s^\delta t^{-1} Z(t)+t^\delta s^{-1} Z(s)\,
  \mu(ds,dt)}{\displaystyle \int_{(0,1]^2} s^\delta +t^\delta\,
  \mu(ds,dt)}.
  \end{equation}
\end{theorem}

\begin{remark} \label{rem:mainrem}
   If $\sup_{0< t\le 1}|R_n(t)|=o\big((nv_n)^{-1/2}\big)$,
    then assertion \eqref{eq:main} holds if merely convergence
    \eqref{eq:blockpreasympest} is required instead of \eqref{eq:blockasympest}, that is,
    the smoothness assumption \eqref{eq:thetantcont} on
    $\theta_{n,t}$ is not needed.
\end{remark}

In \eqref{eq:main} the leading bias term which depends on the
threshold is removed, while the random error is still of the order
$(nv_n)^{-1/2}$. To analyze the latter, w.l.o.g.\ we may assume that
the signed measure $\mu$ is
    symmetric, because $\hat\theta_{n,\mu}=\hat\theta_{n,\tilde\mu}$
    for $\tilde \mu(ds,dt):=\mu(ds,dt)+\mu(dt,ds)$ and $\tilde\mu$ satisfies (M1)--(M3) iff
    $\mu$ meets these conditions. Then the
    right-hand side of \eqref{eq:main} equals
    $$  \frac{\displaystyle \int_{(0,1]^2} s^\delta t^{-1} Z(t)\,
  \mu(ds,dt)}{\displaystyle \int_{(0,1]^2} s^\delta\,
  \mu(ds,dt)}
    $$
    which is a centered Gaussian rv with variance
    $$ \sigma^2_\mu :=
    \frac{\displaystyle \int_{(0,1]^2}  \int_{(0,1]^2} (s\tilde s)^\delta (t\tilde t)^{-1} c(t,\tilde t)\,
  \mu(ds,dt)\,\mu(d\tilde s,d\tilde t)}{\Big(\displaystyle \int_{(0,1]^2} s^\delta\,
  \mu(ds,dt)\Big)^2}.
   $$
   If $\mu$ is the symmetrized version of the signed measure
   discussed in Example \ref{ex:mu} (i) with $f$ and $g$ denoting Lebesgue densities of $Q_F$ and $Q_G:=Q_F^{T_b}$,
   respectively, then
   \begin{eqnarray*}
    \sigma^2_\mu & = &
   \bigg(\frac{ab}{(1-a^{-\delta})(1-b^{-\delta})}\bigg)^2\times\\
   & & \times
   \int_0^1\int_0^1 \Big(
   a^{-(\delta+1)}f(bt)+b^{-(\delta+1)}f(at)-f(abt)-(ab)^{-(\delta+1)}f(t)\Big)\times\\
   & & \hspace*{1cm} \times
   \Big(
   a^{-(\delta+1)}f(b\tilde t)+b^{-(\delta+1)}f(a\tilde t)-f(ab\tilde t)-(ab)^{-(\delta+1)}f(\tilde
   t)\Big)(t\tilde t)^{-1}c(t,\tilde t)\, dt\,d\tilde t.
   \end{eqnarray*}
To estimate this asymptotic variance is essentially as difficult as
to determine the asymptotic variance of the original blocks
estimators. To this end, one may employ ideas developed in Drees
(2003), but a bootstrap approach, that will be worked out in a
forthcoming paper, seems more promising.

Finally, we would like to mention that our approach is obviously not
capable of removing the part $\theta_n-\theta$ of the bias which
does not depend on the threshold but on the block length $r_n$.

\section{Proofs}
\label{sect:proofs}

\begin{proofof}  Theorem \ref{theo:empproc}.\rm \quad
  We apply Theorem 2.10 of Drees and Rootz\'{e}n (2010) to prove
  asymptotic equicontinuity of the processes and Theorem 2.3 to
  establish convergence of the finite dimensional marginal
  distributions. To this end, we must verify the conditions
  required in these theorems.

  (i) The assumptions (B1) and (B2) of Drees and Rootz\'{e}n (2010)
  follow from our conditions (C1) and (C2). For the functionals $f_t$, condition (C2) of Drees and Rootz\'{e}n
  (2010) is trivial.
  Condition (C3) of Drees and Rootz\'{e}n
  (2010) reads as
  $$ \frac{P\{\max_{1\le i\le r_n} U_{n,i}>(1-s)\vee
  (1-t)\}}{r_nv_n} \;\to\; \theta(s\wedge t)
  $$
  (cf.\ Drees and Rootz\'{e}n (2010), (4.1)). This is immediate from
  \eqref{eq:recimean}, which implies
  \begin{equation}  \label{eq:recimeannew}
  \frac{P\{\max_{1\le i\le r_n}U_{n,i}>1-t\}}{r_n v_n t} \;\to\; \theta
  \end{equation}
  uniformly for $t\in (0,1]$.

  Likewise, condition (D3) of Drees and Rootz\'{e}n (2010) is equivalent
  to
  $$ \lim_{\delta\downarrow 0} \limsup_{n\to\infty} \sup_{0\le s\le t\le 1, t-s\le\delta}
   \frac{P\{1-t<\max_{1\le i\le r_n} U_{n,i}\le  1-s\}}{r_nv_n}=0,
  $$
  which again is a direct consequence of the uniform convergence \eqref{eq:recimeannew}.

  The remaining conditions can be verified by the arguments given in
  Drees and Rootz\'{e}n (2010), Section 4 and the proof of Corollary 4.3. (Note that
  $Z_n(f_t)$ equals the random variable $\tilde Z_n(1-t)$ defined in
  Example 4.2 (with $k=1$) of that paper.)
   \medskip

  (ii) This assertion is a reformulation of the results on the
  univariate tail empirical process given in Example 3.8 of Drees and Rootz\'{e}n
  (2010).
  \medskip

  (iii) The equicontinuity of the joint process immediately follows
  from the equicontinuity of $(Z_n(f_t))_{0\le t\le 1}$ and $(Z_n(g_t))_{0\le t\le
  1}$ and a similar remark applies to the conditions (C1) and (C2)
  of Drees and Rootz\'{e}n (2010). The remaining condition (C3) follows
  from (C3.1) and (C3.2) of the present paper and the calculations
  in part (i) above.
\end{proofof}

\begin{proofof} Remark 2.2. \quad \rm
  The conditions (C3.1) and (C3.2) follow by similar arguments as in Remark 3.7 (ii) of Drees and Rootz\'{e}n
  (2010) (cf.\ also Corollary 2.4 of that paper). Here we have
  \begin{eqnarray*}
    c_g(s,t) & = & E\Big(
  \Indd{(1-s,1]}(W_1)\Indd{(1-t,1]}(W_1)\\
    & & \hspace*{1cm} { } +\sum_{k=2}^\infty
  \Indd{(1-s,1]}(W_1)\Indd{(1-t,1]}(W_k)+\Indd{(1-t,1]}(W_1)\Indd{(1-s,1]}(W_k)\Big),
  \end{eqnarray*}
  which equals the right hand side of \eqref{eq:cg},
  and
  $$c_{fg}(s,t) = E\Big(
  \Indd{(1-s,1]}(\max_{i\ge 1}W_i)\sum_{k=1}^\infty
  \Indd{(1-t,1]}(W_k)-\Indd{(1-s,1]}(\max_{i\ge 2}W_i)\sum_{k=2}^\infty
  \Indd{(1-t,1]}(W_k)\Big).
  $$
  If $s\ge t$ and the first sum does not vanish, then the first
  indicator equals 1. Together with a similar reasoning for the
  second sum, one obtains
  $$ c_{fg}(s,t)=E(\Indd{(1-t,1]}(W_1))=t. $$
  In the case $s<t$, direct calculations show that
  \begin{eqnarray*}
    c_{fg}(s,t) & = & E\Big(
  \Indd{(1-s,1]}(\max_{i\ge 1}W_i)  \Indd{(1-t,1]}(W_1)\\
  & & \hspace*{1cm} { } +\big(\Indd{(1-s,1]}(\max_{i\ge 1}W_i)-\Indd{(1-s,1]}(\max_{i\ge 2}W_i)\big)
  \sum_{k=2}^\infty  \Indd{(1-t,1]}(W_k)\Big)
  \end{eqnarray*}
  is equal to the right hand side of \eqref{eq:cfg}.
\end{proofof}

\begin{proofof} Corollary \ref{cor:blockasymp}. \rm\quad
  Using $E(g_t(Y_{n,1}))=r_nv_nt$ and representation
  \eqref{eq:empprocrep}, we obtain by simple calculations
  \begin{equation} \label{eq:errorexpan}
   \sqrt{nv_n}t(\hat\theta_{n,t}^*-\theta_{n,t}) =
  \frac{n}{m_nr_n} \cdot \frac{Z_n(f_t)-\theta_{n,t}Z_n(g_t)}{1+
  \sqrt{nv_n}/(m_nr_nv_nt) Z_n(g_t)}.
  \end{equation}
  By the
  equicontinuity of $(Z_n(g_t))_{0\le t\le 1}$ and $Z_n(g_0)=0$,
  there exists a sequence $\eta_n\to 0$ such that $\sup_{0\le t\le (nv_n)^{-1/4}}
  |Z_n(g_t)|=O_P(\eta_n)$. Hence, for $t_n:=(\eta_n/(nv_n))^{1/2}$,
  $$ \sup_{t_n\le t\le 1} \frac{|Z_n(g_t)|}{\sqrt{nv_n}t}
  =O_P\Big(\frac{\eta_n}{\sqrt{nv_n}t_n}\Big)+\sup_{(nv_n)^{-1/4}\le
  t\le 1}  \frac{|Z_n(g_t)|}{(nv_n)^{1/4}}=o_P(1),
  $$
  so that the denominator of the second fraction tends to 1 uniformly for
  $t\in[t_n,1]$. Moreover, since both $\hat\theta_{n,t}^*$ and
  $\theta_{n,t}$ are bounded,
  \begin{equation}  \label{eq:thetabd1}
    \sup_{0\le t\le t_n}
    \sqrt{nv_n}t|\hat\theta_{n,t}^*-\theta_{n,t}|=o_P(1).
  \end{equation}
  Finally, the continuity of $Z$ implies
  \begin{equation}  \label{eq:Zbd1}
    \sup_{0\le t\le t_n}    |Z(t)|=o_P(1).
  \end{equation}
  Therefore, in view of \eqref{eq:errorexpan}--\eqref{eq:Zbd1},
  Theorem \ref{theo:empproc} and the uniform convergence of
  $\theta_{n,t}$ to $\theta$ prove the assertion.
\end{proofof}

\begin{proofof} Corollary \ref{cor:blockasympest}.\rm\quad
  Check that under the conditions of Theorem \ref{theo:empproc} (iii) the following equivalences
  hold on a set with probability tending to 1: $X_i>X_{n-\lceil nv_nt\rceil:n} \iff U_i>U_{n-\lceil nv_nt\rceil:n}
  \iff U_{n,i} > 1-(1-U_{n-\lceil nv_nt\rceil:n})/v_n$, and thus
  $\hat\theta_{n,t}=\hat\theta^*_{n,s_n(t)}$ with $s_n(t) := (1-U_{n-\lceil
  nv_nt\rceil:n})/v_n$.
  An application of Vervaat's (1972) Theorem 1 to the assertion of Theorem \ref{theo:empproc} (ii) yields
  \begin{equation}  \label{eq:tailempproc}
   \sqrt{nv_n} (s_n(t)-t)_{0\le t\le 1} \;\longrightarrow\; Z_g
  \end{equation}
  (cf.\ the proof of Corollary 3.1 of  Drees (2000)).
  In particular, $s_n(t)/t\to 1$ uniformly for all
  $t\in[(nv_n)^{-1/3},1]$.

  Moreover,  by continuity $\sup_{0\le t\le (nv_n)^{-1/3}}
  |Z(t)|\to 0$ and thus by Corollary \ref{cor:blockasymp}
  \begin{eqnarray}
    \sqrt{nv_n}t\big(\hat\theta_{n,t}-\theta_{n,s_n(t)}\big)1_{[(nv_n)^{-1/3},1]}(t) & = &
  \sqrt{nv_n}s_n(t)\big(\hat\theta_{n,s_n(t)}^*-\theta_{n,s_n(t)}\big)\cdot \frac{t}{s_n(t)}1_{[(nv_n)^{-1/3},1]}(t)
  \nonumber\\
  &\to& Z(t) \label{eq:thetahatconv1}
  \end{eqnarray}
  uniformly for $t\in[0,1]$.

  Next note that
  \begin{equation} \label{eq:thetahatconv2}
    \sup_{0\le t\le (nv_n)^{-3/4}} \sqrt{nv_n} t
    |\hat\theta_{n,t}-\theta_{n,s_n(t)}| \le (nv_n)^{-1/4} \;\longrightarrow\;
    0,
  \end{equation}
  while for $(nv_n)^{-3/4}\le t\le (nv_n)^{-1/3}$
  \begin{eqnarray}
    \lefteqn{\sqrt{nv_n} |\hat\theta_{n,t}-\theta_{n,s_n(t)}|} \nonumber \\
     & = & \Big|\frac{nv_nt}{\lceil nv_nt\rceil}\Big(
     Z_n(f_{s_n(t)}) + \frac{m_n}{\sqrt{nv_n}} P\big\{\max_{1\le
     i\le r_n} X_i>F^\leftarrow(1-v_ns)\big\}|_{s=s_n(t)}\Big) -
     \sqrt{nv_n} t\theta_{n,s_n(t)}\Big| \nonumber \\
     & \le & | Z_n(f_{s_n(t)})| + \Big|\frac{nv_nt}{\lceil
     nv_nt\rceil} \cdot \frac{m_nr_n}n -1\Big| \sqrt{nv_n}  s_n(t) \theta_{n,s_n(t)}
     + \sqrt{nv_n}  |s_n(t)-t| \theta_{n,s_n(t)}.
     \label{eq:thetahatconv3}
  \end{eqnarray}
  The first term on the right-hand side tends to 0 uniformly by
  Theorem \ref{theo:empproc} (i) and the continuity of $Z_f$, the last
  term converges to 0 by \eqref{eq:tailempproc} and the continuity of $Z_g$.
  Furthermore, by \eqref{eq:tailempproc}
  $$ \sup_{(nv_n)^{-3/4} \le t\le (nv_n)^{-1/3}} \Big|\frac{nv_nt}{\lceil
     nv_nt\rceil} \cdot \frac{m_nr_n}n -1\Big| \sqrt{nv_n}  s_n(t) =
     O_P\big( (nv_n)^{-1/4}+ r_n/n\big)\cdot
     O_P\big((nv_n)^{1/6}\big) \to 0.
  $$
  Combining this with
  \eqref{eq:thetahatconv1}--\eqref{eq:thetahatconv3}, we arrive at
  the first assertion.

  It remains to prove that under the additional continuity condition
  on $\theta_{n,t}$
  $$ \sqrt{nv_n}\sup_{0\le t\le
  1}t|\theta_{n,s_n(t)}-\theta_{n,t}|\;\longrightarrow\; 0
  $$
  in probability.
  To this end, first check that
  \begin{eqnarray*}
    |\theta_{n,s}-\theta_{n,t} | & \le & \Big|\frac 1{r_nv_n s}-\frac 1{r_nv_n
    t}\Big| P\big\{\max_{1\le i\le r_n} X_i>F^\leftarrow(1-v_ns)\big\}\\
    & & \hspace{2cm} +
      \frac 1{r_nv_nt}  P\big\{ F^\leftarrow(1-v_n(s\vee t))< \max_{1\le i\le r_n} X_i\le F^\leftarrow(1-v_n(s\wedge
      t))\big\}\\
      & \le & \frac{|t-s|}{r_nv_n st}\cdot r_nv_n s +  \frac
      1{r_nv_nt}\cdot r_nv_n|t-s|\\
      & \le & 2\frac{|t-s|}t.
  \end{eqnarray*}
  Hence, again by \eqref{eq:tailempproc}  and the continuity of
  $Z_g$ for each $\delta>0$ there exists $\eta>0$ such that
  $$ P\big\{ \sqrt{nv_n}\sup_{0\le t\le
  \eta}t|\theta_{n,s_n(t)}-\theta_{n,t}|>\delta\big\}<\delta.
  $$
  On the other hand, by \eqref{eq:tailempproc}, assumption
  \eqref{eq:thetantcont} and Hsing's result \eqref{eq:unifconv}
  $$
   \sqrt{nv_n}t|\theta_{n,s_n(t)}-\theta_{n,t}| =
   \sqrt{nv_n} t \Big|\frac{\theta_{n,s_n(t)}-\theta}{\theta_{n,t}-\theta}-1\Big|\cdot
  |\theta_{n,t}-\theta| =
  O_P\big(|\theta_{n,t}-\theta|\big)\to 0
   $$
uniformly for $t\in[\eta,1]$, which completes the proof.
\end{proofof}

\begin{proofof} Theorem \ref{theo:main}.\rm\quad
  By condition (M1)
    \begin{equation}  \label{eq:intvanish}
      \int_{(0,1]^2} (st)^\delta\, \mu(ds,dt)=0, \qquad
      \mu((0,1]^2)=0.
    \end{equation}
    Thus
    \begin{eqnarray*}
      \hat\theta_{n,\mu}-\theta_n & = & \frac{\displaystyle \int
      (\hat\theta_{n,s}-\theta_n)(\hat\theta_{n,t}-\theta_n)\,
      \mu(ds,dt)}{\displaystyle \int
      (\hat\theta_{n,s}-\theta_n)+(\hat\theta_{n,t}-\theta_n)\,
      \mu(ds,dt)}\\
      & = & \frac{\displaystyle \int
      \big(\hat\theta_{n,s}-\theta_{n,s}+c_n s^\delta+R_n(s)\big)\big(\hat\theta_{n,t}-\theta_{n,t}
      +c_nt^\delta+R_n(t)\big)\,
      \mu(ds,dt)}{\displaystyle \int
      \hat\theta_{n,s}-\theta_{n,s}+\hat\theta_{n,t}-\theta_{n,t}+c_n(s^\delta+t^\delta)+R_n(s)+R_n(t)\,
      \mu(ds,dt)}.
    \end{eqnarray*}
    In view of \eqref{eq:blockasympest} and the integrability condition (M3), the right-hand side has the
    same distribution as
    $$ \frac{\displaystyle \int \big((nv_n)^{-1/2} s^{-1}(Z(s)+o_P(1))+c_ns^\delta+R_n(s)\big)
    \big((nv_n)^{-1/2} t^{-1}(Z(t)+o_P(1))+c_nt^\delta+R_n(t)\big)\,
      \mu(ds,dt)}{\displaystyle  c_n\Big(\int
      s^\delta+t^\delta\,\mu(ds,dt)+o_P(1)\Big)+(nv_n)^{-1/2}\Big(\int
      s^{-1}Z(s)+t^{-1} Z(t)\, \mu(ds,dt)+o_P(1)\Big)}.
    $$
    Because of \eqref{eq:intvanish}, the conditions (M2) (M3), $(nv_n)^{-1/2}=o(c_n)$ and $d_n=o(c_n)$ this
    fraction equals
    $$
      \frac{\displaystyle \int (nv_n)^{-1/2} \big(t^\delta s^{-1}
    Z(s)+s^\delta t^{-1}Z(t)\big)+s^\delta R_n(t)+t^\delta R_n(s)
   \,
      \mu(ds,dt)+o_P\big((nv_n)^{-1/2}\big)}{\displaystyle \int
      s^\delta+t^\delta\,\mu(ds,dt)+o_P(1)}.
    $$
    Now the first assertion is obvious and convergence
    \eqref{eq:main} is an immediate consequence of the additional assumption $d_n=o((nv_n)^{-1/2})$
    and the integrability condition (M3).
\end{proofof}

\begin{proofof} Remark \ref{rem:mainrem}. \rm \quad Recall the definition of $s_n(t)$ from the proof of Corollary
\ref{cor:blockasympest}. For $0<\delta\le 1$ and $0<u\le v$ one has
$v^\delta-u^\delta\le v^{\delta-1}(v-u)\le u^{\delta-1}(v-u)$ and
hence $|(s_n(t))^\delta-t^\delta)|\le t^{\delta-1}|s_n(t)-t|$. For
$\delta>1$, the mean value theorem implies $
|(s_n(t))^\delta-t^\delta)|\le\delta |s_n(t)-t|$. Combining both
inequalities with convergence \eqref{eq:tailempproc}, we conclude
$|(s_n(t))^\delta-t^\delta)|=O_P\big( (nv_n)^{-1/2} t^{-1}\big)$.
Moreover,  under the given conditions,
$R_n(s_n(t))=o_P\big((nv_n)^{-1/2} \big)$.
   Hence, $\hat\theta_{n,t}-\theta_n = (nv_n)^{-1/2}
   t^{-1}(Z_n(t)+o_P(1))+c_n t^\delta$ and we proceed as in the proof of
   Theorem \ref{theo:main} to establish \eqref{eq:main}.
\end{proofof}

\vspace{1cm}

{\large \bf References}

\rueck
 Drees, H. (2000).  Weighted approximations of tail processes for $\beta$--mixing random
variables. {\em Ann.\ Appl.\ Probab.} {\bf 10}, 1274--1301.

\rueck
 Drees, H., and Rootz\'{e}n, H. (2010). Limit Theorems for
Empirical Processes of Cluster Functionals. {\em Ann.\ Statist.}
{\bf 38}, 2145--2186.

\rueck
 Ferro, C.A.T., and Segers, J. (2003).
Inference for clusters of extreme values. {\em J.\ Roy.\ Statist.\
Soc.\ B}, {\bf 65}, 545--556.

\rueck
  Hsing, T. (1993). Extremal index estimation for a weakly dependent stationary sequence. {\em
  Ann.\ Statist.} {\bf 21}, 2043--2071.

\rueck
  Robert, C.Y., Segers, J, and Ferro, C.A.T. (2009). A sliding
  blocks estimator for the extremal index. {\em Electron.\ J.\ Stat.} {\bf 3},
  993-–1020.

\rueck
  Rootz\'en, H. (1995). The tail empirical process for
  stationary sequences. Preprint, Chalmers University Gothenburg.

\rueck
  Rootz\'en, H. (2009): Weak convergence of the tail empirical function
  for dependent sequences.  {\em Stoch.\ Proc.\ Appl.} {\bf 119},
  468-490.

\rueck
  Segers, J. (2003). Functionals of clusters of extremes. {\em Adv.\ Appl.\ Probab.} {\bf
  35}, 1028--1045.

\rueck
  Smith, R.L., and Weissman, I. (1994). Estimating the extremal index.
  {\em J.\ Roy.\ Statist.\ Soc.\ B} {\bf 56}, 515--528.

\rueck
  Weissman, I., and Novak, S.Yu. (1998). On blocks and runs estimators of the extremal index.
  {\em J.\ Statist.\ Plann.\ Inference} {\bf 66}, 281--288.

\end{document}